\DeclareUnicodeCharacter{2500}{}
\documentclass[10pt]{article}
\usepackage{xurl} 
\usepackage{cite} 

\usepackage[utf8]{inputenc}
\usepackage[T1]{fontenc}
\usepackage{geometry}
\usepackage{lmodern,microtype}
\usepackage{amsmath,amssymb,amsfonts,amsthm,mathtools}
\usepackage{bm,bbm}
\usepackage{graphicx}
\usepackage{booktabs}
\usepackage{enumitem}
\usepackage{algorithm}
\usepackage{algpseudocode}
\usepackage{hyperref}
\usepackage[nameinlink,capitalize]{cleveref}
\usepackage{tikz-cd}
\usepackage{subcaption} 
\usepackage[font=small,labelfont=bf]{caption}
\usepackage[sort&compress,numbers]{natbib}
\usepackage{xcolor}   
\usepackage{lineno}   
\usepackage{amsmath, amssymb, amsthm, mathrsfs}

\usepackage[utf8]{inputenc}
\usepackage{booktabs}   
\usepackage{multirow}   
\usepackage[table]{xcolor} 
\usepackage{array}      

\definecolor{headerbeige}{RGB}{220, 222, 210}
\definecolor{finetunegreen}{HTML}{B0D7AA} %

\usepackage{amsmath}
\usepackage{amssymb}
\usepackage{algorithm}
\usepackage{algpseudocode}
\usepackage{bm}
\usepackage{booktabs}
\usepackage{caption}

\title{Mathematical Invariant-Enabled Topological Neural Networks for Molecular and Materials Property Prediction}

\author{Yiming Ren$^{1}$, Xiang Liu$^{1}$,  
Mustafa Hajij$^2$, Pietro Li\`{o}$^{3}$, and Guo-Wei Wei$^{1,4,5, *}$
 \\[2mm]
        \\    
 		\\
         $^1$Department of Mathematics,\\
 		  University of Georgia, Athens, GA 30602, USA. \\
         $^2$Department of Data Science, University of San Francisco, CA 94117, USA.\\
         $^3$ Department of Computer Science and Technology, \\ 
         Cambridge University, Cambridge, United Kingdom. \\
        $^4$Department of Biochemistry and Molecular Biology,\\
 		University of Georgia, Athens, GA 30602, USA.   \\
         $^5$School of Computing,\\
		University of Georgia, Athens, GA 30602, USA. 
 	} 

\date{}
\begin{document}

\maketitle

\begingroup
\renewcommand{\thefootnote}{\fnsymbol{footnote}}
\footnotetext[1]{Corresponding author. Email: guowei.wei@uga.edu}
\endgroup

\begin{abstract}

Existing molecular and materials learning approaches often rely on a limited set of structural representations, which may capture only selected aspects of complex three-dimensional structure. Here, we introduce mathematical invariant-enabled topological neural networks (MITNNs), a framework that represents complex structures through multiple complementary mathematical views and integrates them with topological neural architectures. MITNNs combine multiscale invariants from topology, spectral theory, commutative algebra, differential geometry, and discrete curvature, capturing complementary structural information from the same system. Systematic invariant-subset, architecture-subset, and ensemble analyses show that predictive performance depends on how mathematical representations and neural architectures are paired, with selected combinations outperforming individual models and the aggregation of all available components. Across protein--ligand binding, metal--organic framework properties, mutation-induced protein solubility, and molecular toxicity prediction, MITNN consistently outperforms existing methods. These results establish MITNN as a mathematically multimodal framework for scientific machine learning.

\end{abstract}

\newpage

\section{Introduction}
Accurate prediction of molecular and materials properties underpins a broad range of scientific and technological applications, including drug discovery, protein engineering, porous-material design and predictive toxicology \cite{butler2018machine,vamathevan2019applications,jablonka2020big}. Despite rapid advances in machine learning, a fundamental challenge remains: how to represent complex molecular and materials structures in a manner that faithfully captures their three-dimensional (3D) organization, stereochemistry and higher-order relationships across multiple spatial scales. These structural characteristics span local chemical environments and interatomic interactions, mesoscale connectivity and global 3D organization, and are not fully captured by many existing molecular representations \cite{desgagne2026stereopep}. The resulting representation gap can limit the ability of learning models to distinguish structurally distinct systems that exhibit similar low-order descriptors.

Existing approaches make different trade-offs between structural fidelity, invariance and computational complexity. Conventional fixed-length molecular descriptors provide compact representations but may omit spatial relationships that are not explicitly encoded during descriptor construction \cite{bartok2013representing}. Coordinate-aware and geometric graph models preserve more explicit 3D information, but their architectures must account for the invariance or equivariance of molecular systems under atom permutations, translations and rotations \cite{schutt2017schnet,satorras2021n,fuchs2020se}. Moreover, conventional message-passing schemes primarily propagate information through pairwise connectivity and therefore may have limited capacity to represent angular, many-body and higher-order organization unless these relationships are explicitly incorporated into the representation or architecture \cite{gilmer2017neural,gasteiger2020directional,batatia2022mace}. These limitations point to a broader need for representations that encode complementary structural information across topology, geometry, algebra and higher-order organization while respecting the mathematical symmetries inherent to molecular and materials systems.

Mathematical invariants offer a principled route to such representations. By construction, invariants characterize structural properties that remain unchanged under specified transformations, while different mathematical formalisms can capture complementary aspects of a system. Topological data analysis, for example, uses persistent homology (PH) to track the birth and death of connected components, loops and cavities across a filtration, providing scale-dependent summaries of topological organization \cite{edelsbrunner2002topological,zomorodian2004computing,cang2018representability}. Persistent combinatorial Laplacians extend this description by incorporating spectral information associated with the underlying topological structures \cite{wang2020persistent}, while related persistent operators provide representations for labeled, directed and higher-order structures \cite{wei2025persistent}. Differential-geometric approaches characterize local shape and interaction geometry through curvature descriptors \cite{nguyen2019dg}, and Forman persistent Ricci curvature captures the multiscale evolution of discrete curvature in combinatorial structures \cite{wee2021forman}. Persistent commutative algebra provides yet another perspective by encoding algebraic, combinatorial and graded topological information across a filtration \cite{suwayyid2025persistent}. Collectively, these constructions provide mathematically distinct but potentially complementary descriptions of molecular and materials structure, including topological, spectral, algebraic--combinatorial, geometric and curvature information.

Extracting mathematical invariants, however, addresses only one part of the representation problem. An equally important question is how the resulting information should be processed by a learning architecture. One approach is descriptor-based learning, in which mathematical outputs are transformed into fixed-length vectors, images, curves or multiscale arrays and subsequently processed by conventional neural networks \cite{cang2017topologynet,cang2018integration}. Fully connected artificial neural networks (ANNs) can learn nonlinear interactions among global descriptors, whereas convolutional neural networks (CNNs) exploit local patterns in spatially or multiscale-organized representations. A complementary approach incorporates structural organization directly into neural computation. Graph neural networks propagate information through pairwise connectivity, while geometric and equivariant architectures explicitly encode 3D geometry and prescribed transformation behavior \cite{gilmer2017neural,satorras2021n,fuchs2020se}. More recently, higher-order neural architectures have extended message passing beyond pairwise relationships by organizing computation over simplicial complexes, cell complexes, sheaves and copresheaves \cite{bodnar2022neural,hajij2022topological,hajij2023combinatorial,gillespie2024bayesian,hajij2025copresheaf,papamarkou2024position}. In particular, simplicial neural networks (SNNs) propagate information through incidence relations among simplices \cite{ebli2020simplicial}, whereas copresheaf transformer neural networks (CTNNs) organize information exchange through copresheaf-based structure maps \cite{hajij2025copresheaf,liu2026dtdl}. These architectures provide a means of incorporating higher-order organization directly into the learning process rather than encoding it solely through pairwise descriptors.

The diversity of mathematical invariants and neural architectures raises a fundamental modeling question: can complex scientific structures be learned more effectively by combining multiple complementary mathematical views, and how does the utility of each view depend on the neural architecture used to process it? Existing studies have demonstrated the predictive value of individual mathematical invariants, including persistent topological and curvature-based descriptors, and have combined topological information with conventional molecular descriptors or structure-aware neural networks \cite{cang2017topologynet,wee2021forman,verma2024topological}. However, these approaches have generally focused on individual mathematical representations or particular invariant--architecture combinations rather than systematically integrating multiple mathematical formalisms within a common learning framework. Consequently, it remains unclear whether different mathematical invariants provide complementary information, whether higher-order neural architectures can better exploit such information, and whether combining increasingly diverse mathematical representations necessarily leads to improved prediction.

Here we introduce mathematical invariant-enabled topological neural networks (MITNNs), a mathematically multimodal framework that represents molecular and materials structures through multiple complementary mathematical views and integrates them with neural architectures. MITNNs consider five families of mathematical invariants: persistent homology (PH), persistent Laplacian (PL), commutative algebra (CA), element interactive curvature (EIC) and Forman persistent Ricci curvature (FPRC). These invariants characterize complementary aspects of molecular and materials structures through topological, spectral, algebraic--combinatorial, geometric and discrete-curvature information. On the learning side, MITNNs combine these invariant representations with four neural architectures spanning conventional and higher-order learning: artificial neural networks (ANNs), convolutional neural networks (CNNs), simplicial neural networks (SNNs) and copresheaf transformer neural networks (CTNNs). Pairing the five invariant families with the four architectures produces a systematic library of invariant--architecture models within a common experimental framework. Rather than assuming that a particular invariant or neural architecture is universally optimal, MITNNs enable their predictive roles and interactions to be evaluated systematically. We benchmark this framework on five datasets spanning four representative molecular and materials prediction tasks: protein--ligand binding affinity, metal--organic framework (MOF) gas uptake, mutation-induced changes in protein solubility and quantitative toxicity prediction. Across these tasks, MITNNs consistently outperform existing state-of-the-art methods. These results demonstrate the effectiveness of mathematically multimodal learning and highlight the value of integrating complementary mathematical representations with topological neural computation.

\section{Results}

\begin{figure}[htp]
    \centering
    \includegraphics[
        width=\textwidth,
        height=0.85\textheight,
        keepaspectratio
    ]{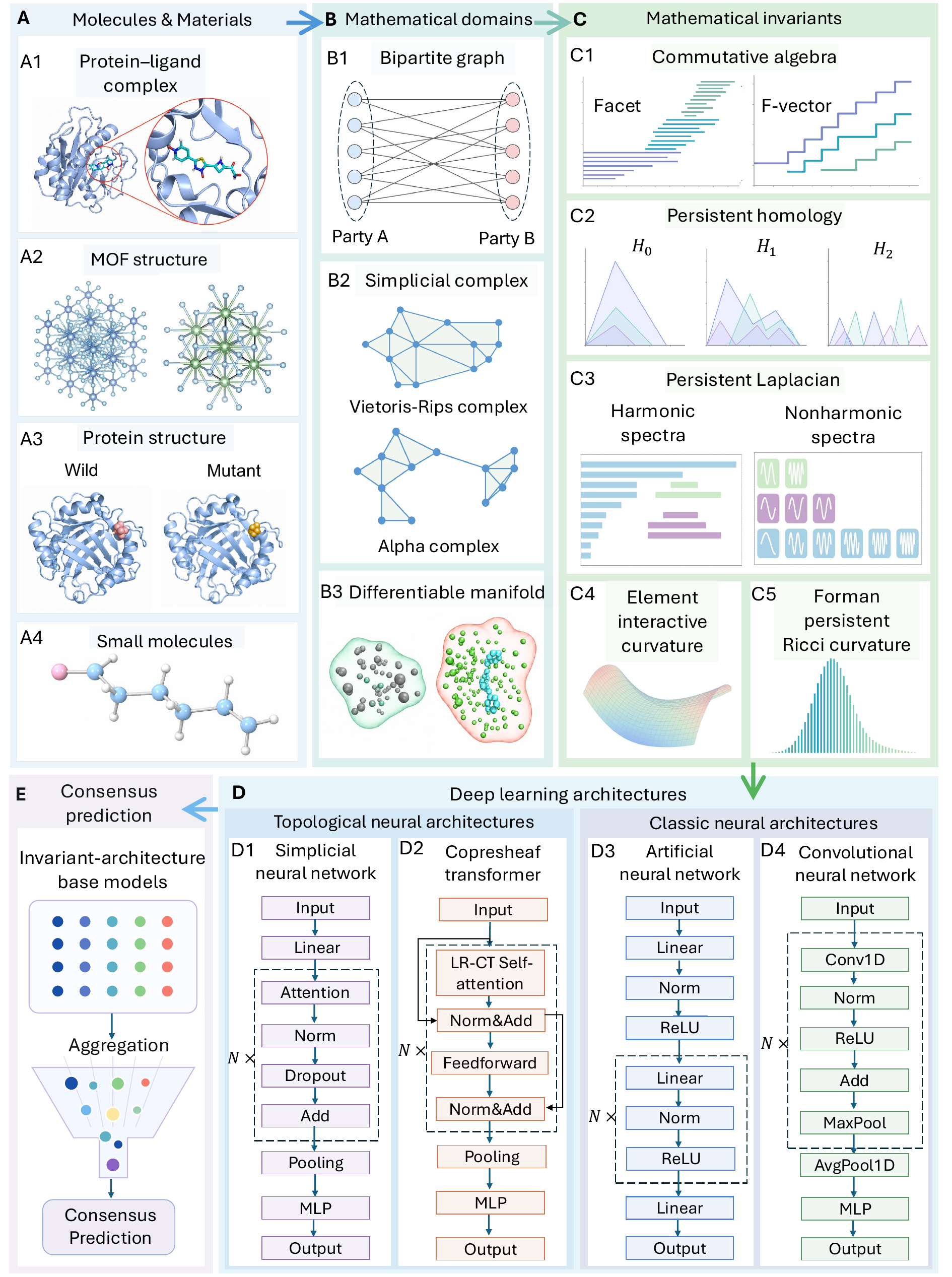}
    \caption{
    {Overview of the MITNN framework.}
    ({A}) Representative molecular and materials systems, including
    ({A1}) protein--ligand complexes,
    ({A2}) metal--organic frameworks,
    ({A3}) wild-type and mutant protein structures, and
    ({A4}) small molecules.
    ({B}) Mathematical domains, including
    ({B1}) bipartite graphs,
    ({B2}) Vietoris--Rips and Alpha simplicial complexes, and
    ({B3}) differentiable manifolds.
    ({C}) Mathematical invariant families, including
    ({C1}) commutative algebra,
    ({C2}) persistent homology,
    ({C3}) persistent Laplacian,
    ({C4}) element interactive curvature, and
    ({C5}) Forman persistent Ricci curvature.
    (D) Neural architectures, including the topological neural architectures (D1) simplicial neural networks and (D2) copresheaf transformer neural networks, and the classic neural architectures (D3) artificial neural networks and (D4) convolutional neural networks.
    ({E}) Consensus prediction obtained by aggregating
   invariant--architecture base models.
    }
    \label{fig:mint_dnn_overview}
\end{figure}

\subsection{Overview of the MITNN Framework}

As illustrated in Fig.~\ref{fig:mint_dnn_overview}, MITNN provides a
unified framework that maps three-dimensional molecular and materials
structures to mathematical domains, constructs multiscale
mathematical invariants, processes the resulting invariant features using
multiple neural architectures, and integrates base-model predictions to form
consensus predictions. The framework accommodates structurally distinct
systems, including protein--ligand complexes, metal--organic frameworks
(MOFs), wild-type and mutant proteins, and small molecules
(Fig.~\ref{fig:mint_dnn_overview}A). Because these systems differ substantially in their structural organization,
MITNN adapts the mathematical domains and invariant constructions to the
characteristics of each application.

The first stage maps three-dimensional molecular and materials structures
to mathematical domains for invariant construction
(Fig.~\ref{fig:mint_dnn_overview}B). Bipartite graphs encode interactions between element-specific atom groups, providing a natural representation of heterogeneous intermolecular interactions. Simplicial complexes encode
proximity and higher-order connectivity across multiple scales, with Vietoris--Rips and alpha complexes providing
distinct constructions for different invariant formulations and
structural settings. Differentiable manifolds provide continuous geometric representations from which local and multiscale geometric quantities can be derived.

From these mathematical domains, MITNN constructs five families of
multiscale mathematical invariants
(Fig.~\ref{fig:mint_dnn_overview}C). Commutative algebra (CA) encodes
algebraic--combinatorial structure through facet counts and $f$-vector
quantities. Persistent homology (PH) tracks the evolution of topological
features across filtration scales in dimensions $H_0$, $H_1$, and $H_2$.
Persistent Laplacian (PL) enriches this multiscale topological information
with spectral characteristics derived from its zero and nonzero eigenvalue
components \cite{wang2020persistent}. Element interactive curvature (EIC)
captures continuous local geometry through curvature quantities derived from
element-resolved interaction manifolds. Forman persistent Ricci curvature
(FPRC) describes the evolution of discrete curvature across filtered
simplicial structures \cite{wee2021forman}. The detailed construction of each mathematical invariant is provided in Section S3.

The resulting invariant features are processed using four neural
architectures (Fig.~\ref{fig:mint_dnn_overview}D), including two
topological neural architectures and two classic neural architectures.
The topological architectures comprise simplicial neural networks (SNNs),
which organize feature propagation according to simplicial relations, and
copresheaf transformer neural networks (CTNNs), which organize information
exchange through copresheaf structure maps and low-rank copresheaf
transformer (LR-CT) self-attention. The LR-CT formulation preserves
structured directional information flow across local feature spaces while
improving computational scalability \cite{liu2026dtdl}. MITNN also
incorporates ANNs for fully connected nonlinear
processing of invariant features and CNNs
for capturing local dependencies in ordered multiscale feature arrays.
Detailed architecture implementations are provided in Section~S4.

Pairing an available mathematical invariant with a neural architecture
produces an invariant--architecture base model. With five invariant families
and four neural architectures, the general MITNN framework admits up to
20 such base models. MITNN supports two consensus variants
(Fig.~\ref{fig:mint_dnn_overview}E).
MITNN$^{\mathrm{all}}$ aggregates all available
invariant--architecture base models, whereas MITNN$^{\mathrm{best}}$
uses the best-performing base-model subset identified through exhaustive
ensemble analysis. Details of the
consensus aggregation procedures for each application and the base-model
compositions of MITNN$^{\mathrm{best}}$ are provided in Section~S2.1 and
Table~S2, respectively.

\begin{figure}[htp]
    \centering
    \includegraphics[width=\linewidth]{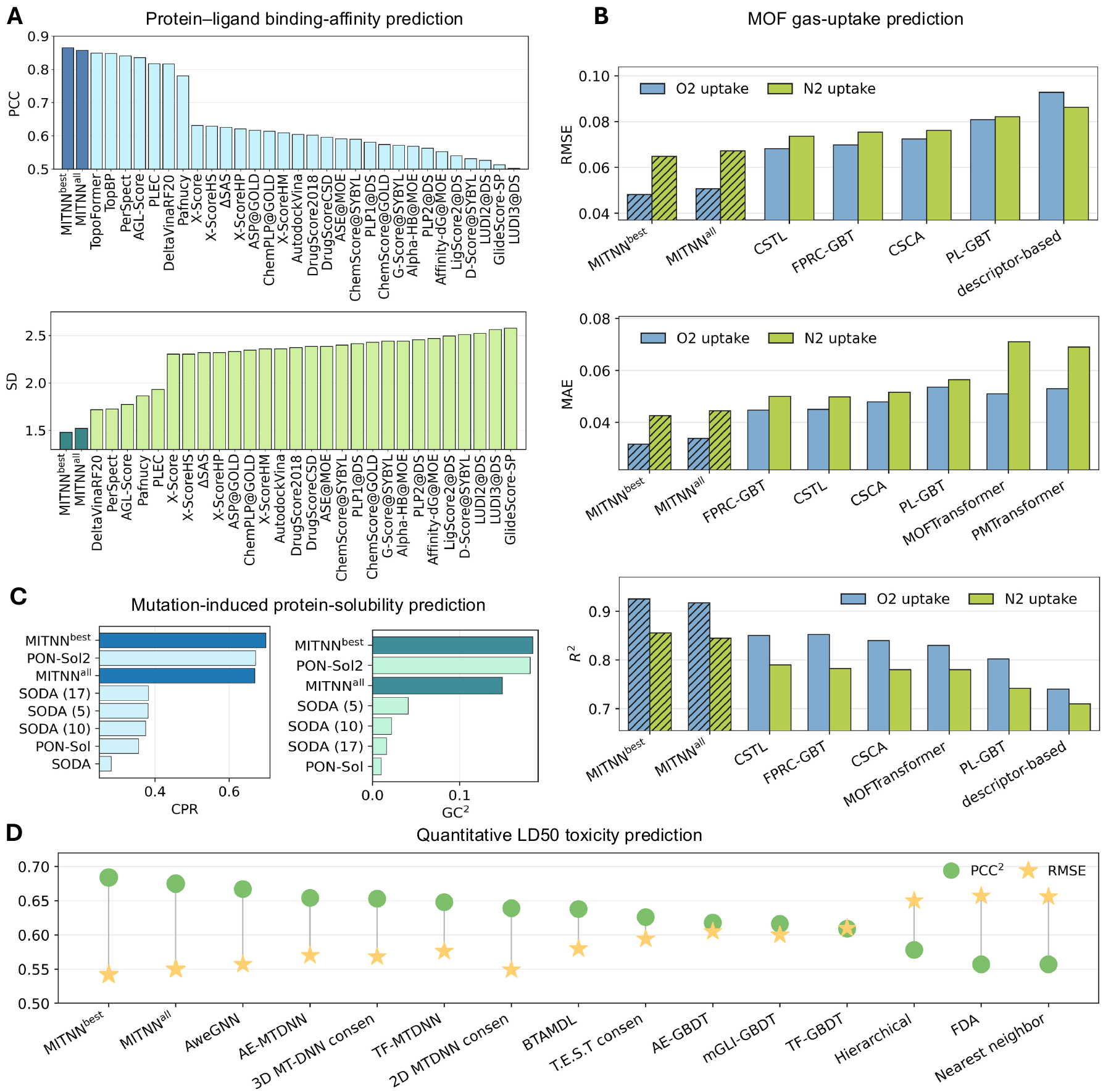}
    \caption{
    Overall predictive performance of MITNN across molecular and materials
    benchmark applications. Results are shown for both MITNN consensus
variants.
    (A) Protein--ligand binding affinity prediction on CASF-2016, evaluated
    by PCC and SD and compared with representative scoring functions and
    topology-based models
    \cite{stepniewska2018development,cang2018representability,
    meng2021persistent,li2014comparative,su2018comparative,
    cheng2009comparative,wang2017improving,nguyen2019agl,
    wojcikowski2019development}. 
    (B) O$_2$ and N$_2$ uptake prediction in MOFs, evaluated by $R^2$, MAE,
    and RMSE and compared with representative descriptor-based, topology-based,
    and transformer-based approaches
    \cite{kang2023multi,park2023enhancing,orhan2021prediction,
    chen2025category,khaemba2026commutative}.
    (C) Mutation-induced protein solubility prediction on PON-Sol2, evaluated
    by CPR and $\mathrm{GC}^{2}$ and compared with existing solubility
    predictors \cite{yang2021pon,yang2016pon}.
    (D) Quantitative LD$_{50}$ toxicity prediction, evaluated by
    $\mathrm{PCC}^{2}$ and RMSE and compared with representative traditional
    machine-learning, multitask-learning, and molecular-descriptor-based
    methods
    \cite{martin2016user,gao20202d,wu2018quantitative,
    jiang2020boosting,szocinski2021awegnn,moe2020development}.
    }
    \label{fig:results1}
\end{figure}

\subsection{MITNN for Protein--Ligand Binding Affinity Prediction}

Protein--ligand binding affinity quantifies the interaction strength between
a ligand and its macromolecular target and is a central endpoint in
structure-based drug discovery
\cite{cai2024carsidock,ain2015machine}. We evaluated MITNN on the
CASF-2016 benchmark \cite{liu2015pdb}, which has been widely used to assess
the scoring performance of computational binding affinity models
\cite{ballester2010machine,wang2017improving,Li2023Development}.

The evaluation protocol used PDBbind-v2016 refined complexes for model
training and CASF-2016 core complexes for benchmark evaluation. Each
invariant--architecture base model was trained in ten independent runs with
different random seeds to account for stochastic variation in model
initialization and training. Predictions from the ten runs were summarized
by their median for each base model, and the MITNN consensus predictions were
then obtained by taking the median across the included base models.
Predictive performance was evaluated using the Pearson correlation
coefficient (PCC) and standard deviation (SD), as defined in
Section~S1.2. Binding affinities originally expressed on the $pK_d$ scale
were converted to kcal/mol by multiplying both experimental and predicted
values by a factor of $-1.36$.

As shown in Fig.~\ref{fig:results1}A, MITNN$^{\mathrm{all}}$ achieved a PCC of 0.857 and an SD of 1.522~kcal/mol, while MITNN$^{\mathrm{best}}$ further improved the PCC to 0.865 and reduced the SD to 1.483~kcal/mol. As summarized in Table~S3, both consensus models outperformed all existing scoring functions and topology-based learning methods \cite{stepniewska2018development,cang2018representability,
    meng2021persistent,li2014comparative,su2018comparative,
    cheng2009comparative,wang2017improving,nguyen2019agl,
    wojcikowski2019development}. In particular, MITNN$^{\mathrm{best}}$ achieved the highest PCC and the lowest reported SD among all evaluated methods. Relative to TopoFormer-Seq, the strongest prior method in the comparison, MITNN$^{\mathrm{best}}$ increased the PCC from 0.852 to 0.865 and reduced the SD from 1.612 to 1.483~kcal/mol, corresponding to a 1.5\% relative improvement in PCC and an 8.0\% reduction in SD. MITNN also outperformed other competitive approaches, including PerSpect-ML \cite{meng2021persistent}, AGL-Score, PLEC, DeltaVinaRF20, and Pafnucy.

\subsection{MITNN for MOF Gas-Uptake Prediction}

Metal--organic frameworks (MOFs) are porous crystalline materials whose gas
adsorption properties depend on both global structural characteristics, such
as pore geometry and framework connectivity, and local chemical environments.
We therefore evaluated MITNN on MOF gas-uptake prediction to assess its
performance on a materials property prediction task involving both global
structural organization and local chemical environments. To account for elemental heterogeneity, the constituent elements were grouped
into eight chemical categories together with an all-atom category
(Fig.~S3).

We evaluated MITNN for O$_2$ and N$_2$ uptake prediction, with uptake
reported in mol/kg. The dataset sizes and partitioning protocols are
summarized in Table~S1. For each gas species, ten random
train--validation--test partitions were generated using an 8:1:1 ratio.
Within each partition, each invariant--architecture base model was evaluated
over multiple independently seeded runs. The predictions from the included
base models and independent runs were averaged to obtain the corresponding
MITNN consensus prediction. Predictive performance was evaluated using
$R^2$, MAE, and RMSE, as defined in Section~S1.2. The reported metrics were
averaged across the ten random partitions.

As shown in Fig.~\ref{fig:results1}B, MITNN$^{\mathrm{all}}$ achieved an
$R^2$ of 0.909 for O$_2$ uptake, which increased to 0.917 for
MITNN$^{\mathrm{best}}$. For N$_2$ uptake, MITNN$^{\mathrm{all}}$ and
MITNN$^{\mathrm{best}}$ achieved $R^2$ values of 0.838 and 0.845,
respectively. MITNN$^{\mathrm{best}}$ also yielded lower MAE and RMSE than MITNN$^{\mathrm{all}}$ for both gas species. Both MITNN consensus models outperformed all representative descriptor-based, topology-based, and transformer-based approaches included in our evaluation. In particular, compared with the strongest non-MITNN baselines, MITNN$^{\mathrm{best}}$ improved $R^2$ by 7.9\% for O$_2$ uptake and 7.0\% for N$_2$ uptake, while reducing MAE by 23.9\% and 15.1\%, and RMSE by 23.3\% and 13.8\%, respectively. These comparisons include MOFTransformer \cite{kang2023multi}, PMTransformer \cite{park2023enhancing}, a descriptor-based model \cite{orhan2021prediction}, CSTL \cite{chen2025category}, CSCA \cite{khaemba2026commutative}, as well as the topology-based FPRC-GBT and PL-GBT baselines evaluated under the same ten-partition protocol. Detailed numerical comparisons are provided
in Table~S4, which also includes PL-GBT and FPRC-GBT, obtained by coupling
persistent Laplacian and Forman persistent Ricci curvature features,
respectively, with gradient-boosting trees.

\subsection{MITNN for Mutation-Induced Protein Solubility Prediction}

Mutation-induced changes in protein solubility can influence protein stability,
aggregation propensity, and biological function. We evaluated MITNN on the
PON-Sol2 benchmark \cite{yang2021pon}, which formulates mutation-induced
solubility prediction as a three-class classification problem comprising
decreased, neutral, and increased solubility. Following the position-stratified blind-test protocol of PON-Sol2
\cite{yang2021pon}, models were evaluated on an independent test set of
662 variants, including 338 solubility-decreasing, 87 neutral, and 237
solubility-increasing mutations. The dataset sizes and partitioning protocol
are summarized in Table~S1. Each invariant--architecture base model was
trained in ten independent runs with different random seeds. Predictions from
the included base models and independent runs were averaged before class
assignment. Predictive performance was evaluated using the Correct Prediction
Ratio (CPR) and Generalized Squared Correlation ($\mathrm{GC}^{2}$), as
defined in Section~S1.2. As shown in Fig.~\ref{fig:results1}C,
MITNN$^{\mathrm{all}}$ achieved a CPR of 0.669 and a
$\mathrm{GC}^{2}$ of 0.149. MITNN$^{\mathrm{best}}$ improved these values
to 0.699 and 0.184, respectively, outperforming the published PON-Sol2
predictor, which reported a CPR of 0.671 and a $\mathrm{GC}^{2}$ of 0.181.
Detailed comparisons with the evaluated protein solubility predictors are
provided in Table~S5.

\subsection{MITNN for Quantitative Toxicity Prediction}

Quantitative toxicity prediction is important for chemical safety assessment
and drug development. We evaluated MITNN on the oral rat median lethal
dose (LD$_{50}$) endpoint, a widely used measure of acute toxicity
\cite{wu2018quantitative}. 

Each invariant--architecture base model was trained in ten independent runs
with different random seeds. Predictions were summarized by the median across
the independent runs for each base model, and the MITNN consensus was then
obtained by taking the median across the included base models. Predictive
performance was evaluated using the squared Pearson correlation coefficient
($\mathrm{PCC}^{2}$) and root mean square error (RMSE), as defined in
Section~S1.2. As shown in Fig.~\ref{fig:results1}D,
MITNN$^{\mathrm{all}}$ achieved a $\mathrm{PCC}^{2}$ of 0.675 and an RMSE of 0.550, whereas MITNN$^{\mathrm{best}}$ further improved these values to 0.684 and 0.542, respectively. As summarized in Table~S6, MITNN$^{\mathrm{best}}$ achieved the highest $\mathrm{PCC}^{2}$ and the lowest RMSE among all evaluated toxicity-prediction methods, outperforming conventional toxicity-estimation approaches, molecular-descriptor-based machine-learning models, multitask-learning methods, and graph-based neural networks \cite{martin2016user,wu2018quantitative,gao20202d,szocinski2021awegnn,moe2020development}.

\begin{figure}[htp]
\centering
\includegraphics[width=1.0\linewidth]{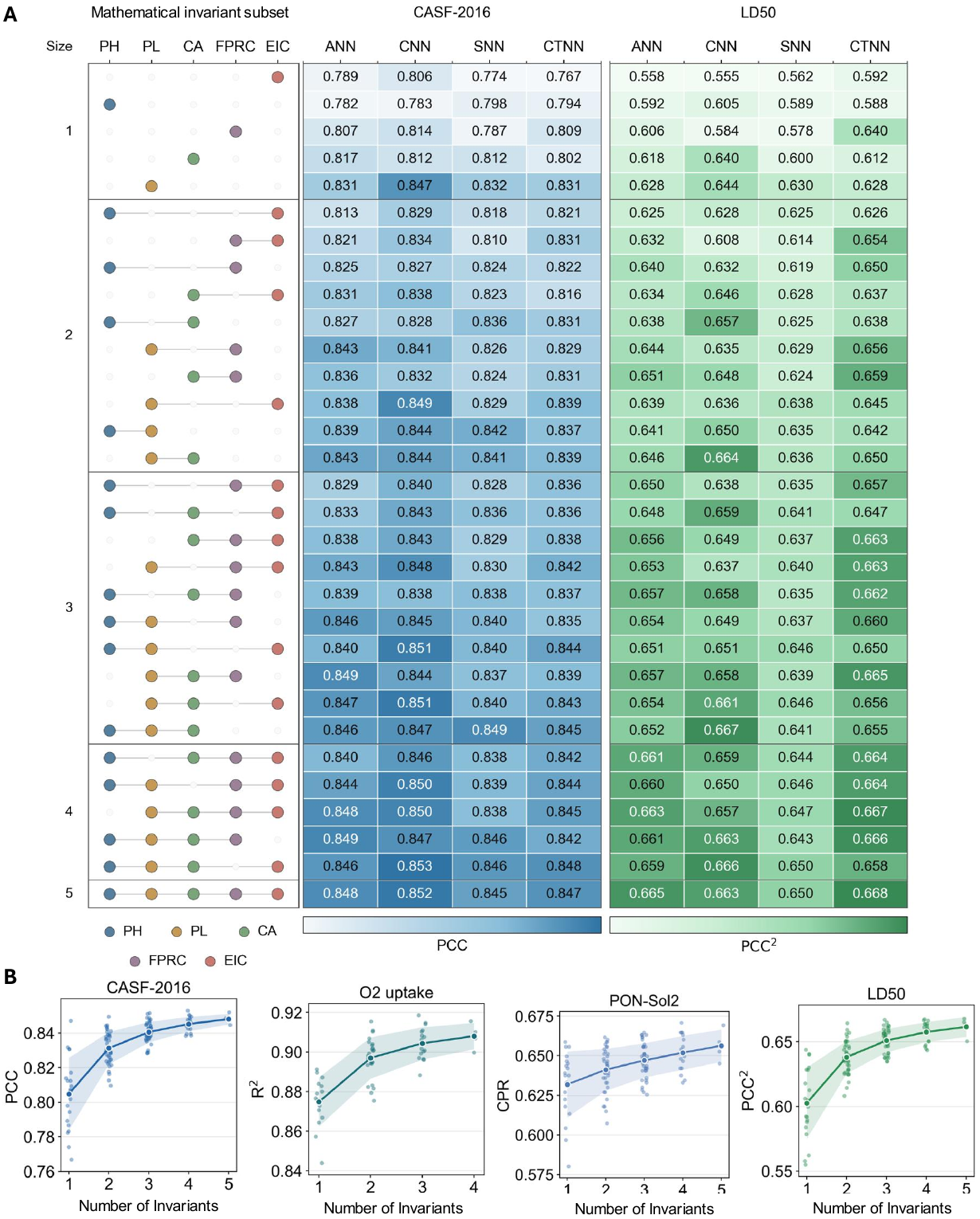}
\vspace{-20pt}
\caption{{Complementarity among mathematical invariant families in MITNN.}
({A}) Exhaustive invariant-subset analysis for CASF-2016 and LD$_{50}$. 
The incidence matrix indicates the invariant families included in each subset. 
For each neural-network architecture, predictions from the corresponding base models were averaged to form the subset prediction. 
The heatmaps report PCC for CASF-2016 and $\mathrm{PCC}^{2}$ for LD$_{50}$ across four architectures.
({B}) Performance versus invariant-subset size for CASF-2016, MOF O$_2$ uptake, PON-Sol2, and LD$_{50}$. 
Each point denotes one invariant subset under a fixed architecture; solid lines show the mean over subsets of the same size across the four architectures, with shaded regions indicating variability. 
CASF-2016, PON-Sol2, and LD$_{50}$ use PH, PL, CA, FPRC, and EIC, while MOF O$_2$ uptake uses PH, PL, CA, and FPRC.}
\label{fig:descriptor_complementarity_ablation}
\end{figure}

\begin{figure}[htp]
\centering
\includegraphics[width=1.0\linewidth]{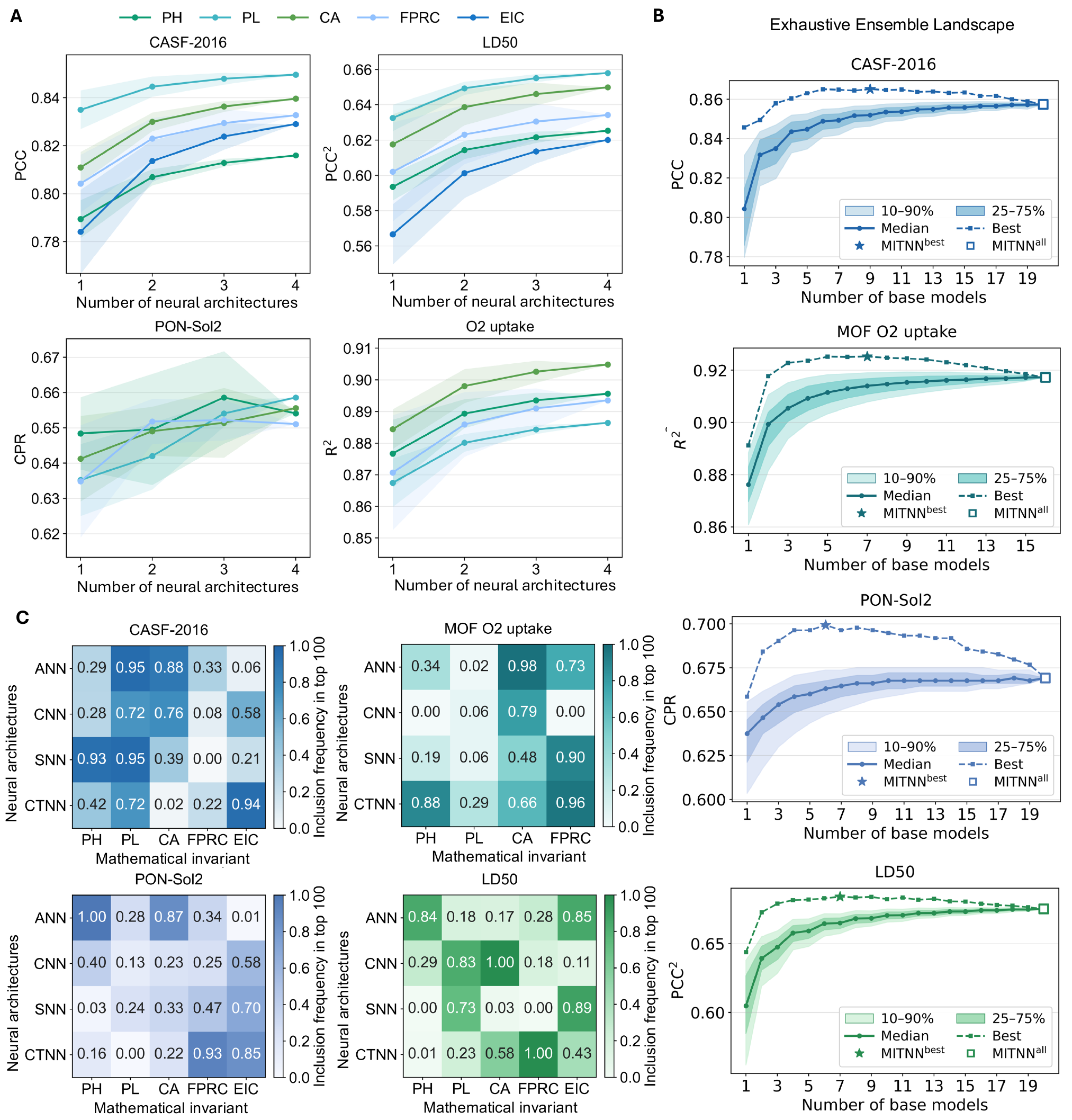}
\caption{{Architectural complementarity and selective ensemble composition
in MITNN.}
({A}) Architecture-subset analysis with the mathematical invariant held
fixed. For each invariant, predictions from the corresponding
invariant--architecture base models within a given architecture subset were
averaged to form the subset prediction. Points show the mean performance
over all architecture subsets of the same cardinality, and shaded regions
indicate the corresponding variability.
({B}) Ensemble-performance landscapes over the complete
invariant--architecture base-model pool for CASF-2016, PON-Sol2,
LD$_{50}$, and MOF O$_2$ uptake. For each ensemble size, the performance
distribution was obtained over all candidate base-model subsets of that
cardinality. Solid and dashed curves denote the median and best performance
at each ensemble size, respectively, and shaded regions indicate the
25--75 and 10--90 percentile intervals. Stars mark
MITNN$^{\mathrm{best}}$, corresponding to the global maximum of each
ensemble landscape, whereas hollow squares mark
MITNN$^{\mathrm{all}}$ at the complete-pool endpoints.
({C}) Inclusion frequency of each invariant--architecture base model among
the top 100 performing ensembles for each application.}
\label{fig:nn_complementarity_ablation}
\end{figure}

\subsection{Mathematical-Invariant Complementarity within Fixed Neural Architectures}
\label{sec:descriptor_complementarity_ablation}

To assess complementarity among mathematical invariant families, we held the
neural architecture fixed and systematically varied the invariant subset.
For CASF-2016, PON-Sol2, and LD$_{50}$, all 31 nonempty subsets of PH, PL,
CA, FPRC, and EIC were evaluated. For the MOF application, all 15 nonempty
subsets of PH, PL, CA, and FPRC were evaluated for both O$_2$ and N$_2$
uptake. For the MOF application, the complementarity analysis presented here focuses
on O$_2$ uptake, while the corresponding N$_2$ uptake results are provided
in Section~S5. For each fixed architecture, predictions from the base models corresponding
to a given invariant subset were averaged at the sample level. Complete
results are provided in Tables~S7--S10 for CASF-2016,
Tables~S11--S14 for PON-Sol2, Tables~S15--S18 and S19--S22 for MOF O$_2$
and N$_2$ uptake, respectively, and Tables~S23--S26 for LD$_{50}$.

Figure~\ref{fig:descriptor_complementarity_ablation}A shows the performance
of all invariant combinations for CASF-2016 and LD$_{50}$ under each fixed
neural architecture, whereas
Figure~\ref{fig:descriptor_complementarity_ablation}B summarizes performance
as a function of invariant-subset size across all four applications.
Overall, combining multiple invariant families tends to improve predictive
performance, with the largest average gain generally occurring from one to
two invariants and smaller gains as additional invariants are included. This trend supports complementarity among mathematically distinct
invariant families.
However, performance also depends strongly on invariant composition. For
PON-Sol2, subsets of the same size show substantial performance variability.
For CASF-2016, CNN achieves its highest PCC of 0.853 with
PH+PL+CA+EIC, slightly exceeding the PCC of 0.852 obtained with all five
invariant families, indicating that the complete invariant set is not
necessarily optimal.

Invariant preferences also vary across applications and neural architectures.
For CASF-2016, PL appears consistently in high-performing combinations across
architectures. For MOF O$_2$ uptake, CA and FPRC occur frequently in
high-performing subsets, with PH or PL providing additional gains depending
on the architecture. For PON-Sol2 and LD$_{50}$, the best-performing
invariant combinations vary more substantially across architectures.
Figure~S4 further summarizes the distributions of invariant-subset
performance under each fixed architecture. CNN shows the highest central
tendency for CASF-2016, whereas CTNN shows the highest central tendency for
MOF O$_2$ uptake and LD$_{50}$; the PON-Sol2 distributions are more similar
across architectures.

\subsection{Architectural Complementarity for Fixed Mathematical Invariants}
\label{sec:nn_complementarity_ablation}

Architectural complementarity was examined by holding the mathematical
invariant fixed and systematically varying the subset of neural
architectures. For each invariant, all 15 nonempty subsets of ANN, CNN,
SNN, and CTNN were evaluated. For this architecture-subset analysis,
predictions from the corresponding invariant--architecture base models were
averaged at the sample level. 
Complete results are provided in Tables~S27--S31 for CASF-2016,
Tables~S32--S36 for PON-Sol2, Tables~S37--S40 and S41--S44 for MOF O$_2$
and N$_2$ uptake, respectively, and Tables~S45--S49 for LD$_{50}$.

Figure~\ref{fig:nn_complementarity_ablation}A summarizes performance as a
function of architecture-subset size for each fixed mathematical invariant.
Across most invariant--application settings, combining multiple architectures
improves average performance relative to using a single architecture, although
the improvement is not necessarily monotonic and depends on architecture
composition. For CASF-2016, PL remains consistently strong across architecture
subsets, indicating relatively low sensitivity to architecture composition.
For MOF O$_2$ uptake, CA achieves the strongest overall performance, whereas
FPRC shows larger gains from combining multiple architectures. For PON-Sol2,
the overlapping and nonmonotonic performance trends indicate that the specific
architecture composition is more influential than subset size alone. For
LD$_{50}$, FPRC performs strongly with CTNN alone, whereas EIC benefits more
substantially from combining multiple architectures. 
\subsection{Selective Integration of Invariant–Architecture Base Models}
\label{sec:component_ensemble_landscape}

The preceding complementarity analyses show that predictive performance
depends jointly on the mathematical invariant and the neural architecture.
We therefore examined ensembles of available invariant--architecture base
models to assess how the number and composition of included base models
influence performance. Candidate subsets were evaluated using the aggregation
procedures defined for each application.

Figure~\ref{fig:nn_complementarity_ablation}B shows the
ensemble-performance landscapes for CASF-2016 binding affinity, MOF O$_2$ uptake,
PON-Sol2 protein solubility, and LD$_{50}$ toxicity. Across all four applications, median performance increases
rapidly for smaller ensembles and then gradually approaches a plateau as
more base models are included. Meanwhile, the 25--75 and 10--90 percentile
intervals become progressively narrower, indicating reduced variability
among base-model subsets of the same size. Despite this overall trend, the
highest performance is reached before all available base models are included.
The best-performing ensembles contain 9, 7, 6, and 7 base models for
CASF-2016, MOF O$_2$ uptake, PON-Sol2, and LD$_{50}$, respectively, with
their corresponding base-model compositions summarized in Table~S2.
Additional base-model-level comparisons are provided in Figs.~S1 and S2.
Overall, increasing ensemble size generally improves typical performance and
reduces variability, whereas optimal performance depends on base-model
composition rather than ensemble size alone.

Figure~\ref{fig:nn_complementarity_ablation}C further characterizes the
composition of high-performing ensembles by reporting the inclusion
frequencies of individual invariant--architecture base models among the top
100 ensembles. The resulting frequency distributions are markedly
nonuniform and vary across applications, indicating that certain base models
are preferentially selected. For CASF-2016, PL-based models appear
frequently across multiple architectures, whereas CA- and FPRC-based models
are more prominent for MOF O$_2$ uptake. These patterns are consistent with
the invariant-level complementarity results in
Section~\ref{sec:descriptor_complementarity_ablation} and broadly agree with
the compositions of MITNN$^{\mathrm{best}}$ summarized in Table~S2.
Overall, high-performing ensembles favor specific invariant--architecture
pairings rather than sampling uniformly from the available base-model pool,
further highlighting the importance of ensemble composition for predictive
performance.

\section{Discussion}

MITNN provides a unified, mathematically multimodal framework for integrating complementary mathematical representations with multiple neural architectures in molecular and materials property prediction. The framework combines five invariant families that capture algebraic-topological, spectral, algebraic--combinatorial, continuous differential-geometric, and discrete-curvature information, providing distinct mathematical views of the same underlying structures. On the learning side, MITNN incorporates topological neural architectures, including simplicial neural networks and copresheaf transformer neural networks, alongside artificial and convolutional neural networks. Coupling the available mathematical invariants with these architectures forms a library of invariant--architecture base models, enabling mathematical and architectural complementarity to be examined within a common predictive framework. Across protein--ligand binding affinity, MOF gas uptake, mutation-induced protein solubility, and quantitative toxicity, MITNN achieves consistently strong performance relative to the evaluated baselines.

The complementarity analyses show that predictive performance depends jointly
on the mathematical invariant and the neural architecture used to process its
features. Combining multiple invariant families or neural architectures is
generally beneficial, but the extent of improvement depends on the specific
invariants, architectures, and prediction task. For example, for CASF-2016,
the frequent representation of PL-based models among high-performing
combinations may reflect the additional spectral information captured by PL
alongside multiscale topological connectivity. 

The ensemble analyses further show that broader aggregation improves
typical performance and reduces variability, whereas the highest performance
is achieved by selected subsets of invariant--architecture base models. The
nonuniform inclusion frequencies among high-performing ensembles further
indicate that certain invariant--architecture pairings are preferentially
represented. These results demonstrate that predictive gains depend not only
on the diversity of mathematical invariants and neural architectures, but
also on how they are paired and integrated.

The conclusions of this study are conditioned on the invariant families,
neural architectures, and benchmark applications considered here, and the observed trends should therefore not be interpreted as universally
applicable across all prediction problems. Future
extensions may incorporate additional mathematical constructions, including
persistent path Laplacians \cite{wang2023persistent}, persistent hyperdigraph
Laplacians \cite{chen2023persistent}, 
and localized persistent commutative algebra \cite{he2026localized},
together with
additional topological neural architectures
\cite{bodnar2022neural,hajij2023combinatorial,gillespie2024bayesian}
and geometric and equivariant neural networks
\cite{satorras2021n,fuchs2020se,batatia2022mace}. As the invariant--architecture library grows, scalable
strategies for base-model screening and selection will also become important. More broadly, MITNN may be extended to other complex scientific data in which complementary mathematical representations of stereochemistry need to be coupled with high-order neural processing mechanisms.

\section{Methods}

\subsection{Mathematical Invariants}
\label{sec:mathematical_invariant_descriptors}

MITNN employs five families of multiscale mathematical invariants:
PH, PL, CA, EIC, and FPRC. In general, let
$D_{\lambda}$ denote a mathematical quantity constructed from a mathematical
domain $\mathcal{X}$ at scale $\lambda$. For an admissible transformation
$g$, $D_{\lambda}$ is invariant with respect to $g$ if
\[
D_{\lambda}(g\mathcal{X}) = D_{\lambda}(\mathcal{X}).
\]
The admissible transformations depend on the particular invariant
construction and may include label-preserving atom or vertex reindexings,
Euclidean rigid motions, and simplicial isomorphisms.

The five invariant families characterize distinct aspects of molecular and
materials structure. PH characterizes the persistence of topological
features across a filtration; PL augments this topological information with
spectral quantities derived from its zero and nonzero eigenvalues; CA encodes
algebraic--combinatorial properties of filtered simplicial complexes; EIC
characterizes continuous geometric information derived from element-specific
interaction density fields; and FPRC describes the evolution of discrete
combinatorial curvature over filtered simplicial complexes. Because these
quantities are evaluated across filtration or interaction scales, their
values generally vary with $\lambda$, while remaining invariant under the
admissible transformations at each fixed scale. The resulting scale-indexed
quantities are organized into multiscale invariant feature blocks and
subsequently supplied to the neural architectures. Further details of invariant construction and feature generation are provided
in Section~S3.

\subsection{Simplicial Complexes and Filtrations}
\label{sec:simplicial_complexes_filtration}

We first introduce the algebraic-topological notation shared by the
filtration-based invariant constructions. A $q$-simplex is the
convex hull of $q+1$ affinely independent vertices.  In particular, $0$-, $1$-, $2$-, and
$3$-simplices correspond to vertices, edges, filled triangles, and
tetrahedra, respectively. A simplicial complex $\mathcal{K}$ is a finite collection of simplices
that is closed under taking faces: if $\sigma\in\mathcal{K}$, then every
face of $\sigma$ also belongs to $\mathcal{K}$.

Let $\Bbbk$ be a field and let $C_q(\mathcal{K};\Bbbk)$ denote the
$q$-chain group generated by the oriented $q$-simplices of $\mathcal{K}$.
The boundary operator
\[
\partial_q:
C_q(\mathcal{K};\Bbbk)
\longrightarrow
C_{q-1}(\mathcal{K};\Bbbk)
\]
acts on an oriented simplex as
\[
\partial_q[u_0,\ldots,u_q]
=
\sum_{\ell=0}^{q}
(-1)^\ell
[u_0,\ldots,\widehat{u_\ell},\ldots,u_q],
\]
where $\widehat{u_\ell}$ denotes omission of the vertex $u_\ell$.
Because
$\partial_q\partial_{q+1}=0$, every boundary is a cycle. The $q$-th
homology group is therefore
\[
H_q(\mathcal{K};\Bbbk)
=
\frac{\ker\partial_q}{\operatorname{im}\partial_{q+1}},
\]
and its dimension
\[
\beta_q
=
\dim_{\Bbbk} H_q(\mathcal{K};\Bbbk)
\]
is the $q$-th Betti number. In the usual geometric interpretation,
$\beta_0$, $\beta_1$, and $\beta_2$ quantify connected components,
independent loops, and enclosed cavities, respectively.

\begin{figure}[htp]
    \centering
    \includegraphics[width=0.8\linewidth]{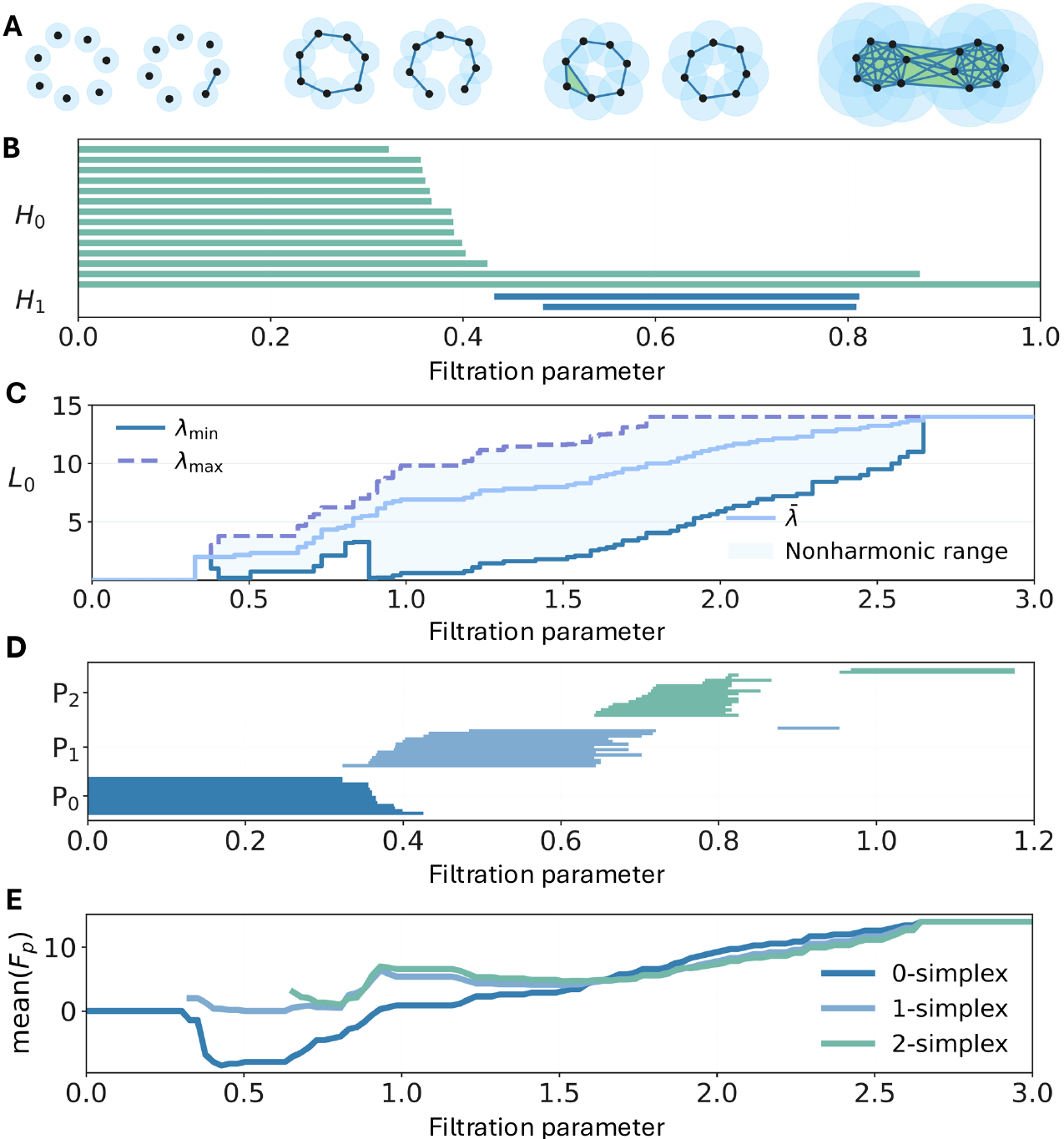}
    \caption{
    Schematic comparison of four filtration-based multiscale mathematical
    invariants obtained from the same Vietoris--Rips filtration of a
    14-point two-loop point cloud.
    (A) Representative Vietoris--Rips complexes at increasing filtration
    parameter $\epsilon$.
    (B) Persistence intervals for $H_0$ and $H_1$. 
    (C) Evolution of the nonharmonic spectrum of the $0$-Laplacian $L_0$,
illustrated by the smallest and largest positive eigenvalues
$\lambda_{\min}^{+}$ and $\lambda_{\max}^{+}$, the mean positive
eigenvalue $\bar{\lambda}$, and the nonharmonic spectral range.
    (D) Facet-persistence bars $P_0$, $P_1$, and $P_2$ associated with
    $0$-, $1$-, and $2$-dimensional facets.
    (E) Signed mean Forman curvature $\operatorname{mean}(F_p)$ for
    $p=0,1,2$.
    }
    \label{fig:math_invariant_example}
\end{figure}

To characterize point-cloud data across geometric scales, we consider a
filtration, namely a nested sequence of simplicial complexes
\[
\mathcal{K}_{t_0}
\subseteq
\mathcal{K}_{t_1}
\subseteq
\cdots
\subseteq
\mathcal{K}_{t_M},
\qquad
t_0<t_1<\cdots<t_M.
\]
Common constructions include Vietoris--Rips and alpha filtrations
~\cite{vietoris1927hoheren,edelsbrunner2010computational,
edelsbrunner2011alpha}.
For $t_a\leq t_b$, the inclusion
\[
\mathcal{K}_{t_a}
\hookrightarrow
\mathcal{K}_{t_b}
\]
captures the evolution of the simplicial structure across scales and
provides the basis for the topological, spectral, algebraic--combinatorial,
and discrete-curvature invariants introduced below.
Figure~\ref{fig:math_invariant_example}A shows representative stages of
such a filtration in the schematic example.

\subsection{Persistent Homology (PH)}
\label{sec:persistent_homology}

Persistent homology (PH) \cite{zomorodian2004computing} extends simplicial homology by tracking the
appearance and disappearance of topological features across a filtration.
For filtration indices $a\leq b$, the inclusion
\[
\mathcal{K}_{t_a}
\hookrightarrow
\mathcal{K}_{t_b}
\]
induces a homomorphism
\[
\phi_{a,b}^{q}:
H_q(\mathcal{K}_{t_a};\Bbbk)
\longrightarrow
H_q(\mathcal{K}_{t_b};\Bbbk).
\]
The corresponding persistent homology group is
\[
H_q^{a,b}
=
\operatorname{im}\phi_{a,b}^{q},
\]
and its dimension
\[
\beta_q^{a,b}
=
\dim_{\Bbbk}H_q^{a,b}
\]
is the persistent Betti number. The birth and death of homological features
across the filtration are summarized by persistence intervals, which provide
a multiscale characterization of topological structure.

PH is invariant under vertex relabeling and filtration-preserving simplicial
isomorphisms. For distance-based filtrations, Euclidean translations and
rotations preserve pairwise distances and therefore leave the persistence
barcodes unchanged. Uniform spatial scaling, however, rescales the filtration
parameters and is not considered an invariant transformation here.

Figure~\ref{fig:math_invariant_example}B illustrates the $H_0$ and $H_1$
persistence intervals for the schematic two-loop point cloud. The $H_0$
intervals describe the progressive merging of connected components as the
filtration expands, whereas the $H_1$ intervals record the formation and
subsequent filling of the two loop-like structures.

\subsection{Persistent Laplacian  (PL)}
\label{sec:persistent_laplacian}

Persistent Laplacian (PL) spectra complement persistent homology by
characterizing both harmonic and nonharmonic spectral properties of
simplicial complexes along a filtration\cite{wang2021hermes}. For the spectral construction,
the chain groups are taken over $\mathbb{R}$ and endowed with the inner
product for which the oriented simplices form an orthonormal basis. The
$q$-th combinatorial Laplacian is
\begin{equation}
    L_q
    =
    \partial_{q+1}\partial_{q+1}^{*}
    +
    \partial_q^{*}\partial_q,
    \label{eq:ordinary_laplacian}
\end{equation}
where $\partial_q^{*}$ denotes the adjoint of the boundary operator.
The operator $L_q$ is real, symmetric, and positive semidefinite, and
induces the discrete Hodge decomposition
\[
    C_q(\mathcal{K};\mathbb{R})
    =
    \operatorname{im}\partial_{q+1}
    \oplus
    \ker L_q
    \oplus
    \operatorname{im}\partial_q^{*}.
\]
Moreover,
\[
    \ker L_q
    \cong
    H_q(\mathcal{K};\mathbb{R}),
\]
so the multiplicity of the zero eigenvalue equals the $q$-th Betti number
~\cite{eckmann1944harmonische}. 
Thus, the zero-eigenvalue multiplicity recovers topological information,
whereas the positive eigenvalues encode additional spectral information
about the incidence and connectivity structure of the simplicial complex.

For filtration indices $a\leq b$, define
\[
    C_{q+1}^{a,b}
    =
    \left\{
    c\in C_{q+1}(\mathcal{K}_{t_b};\mathbb{R})
    \ \middle|\
    \partial_{q+1}^{b}c
    \in C_q(\mathcal{K}_{t_a};\mathbb{R})
    \right\}.
\]
Restricting $\partial_{q+1}^{b}$ to this space gives the persistent boundary
operator
\[
    \partial_{q+1}^{a,b}:
    C_{q+1}^{a,b}
    \rightarrow
    C_q(\mathcal{K}_{t_a};\mathbb{R}),
\]
and the corresponding persistent Laplacian is
\begin{equation}
    L_q^{a,b}
    =
    \partial_{q+1}^{a,b}
    \bigl(\partial_{q+1}^{a,b}\bigr)^{*}
    +
    \bigl(\partial_q^{a}\bigr)^{*}\partial_q^{a}.
    \label{eq:persistent_laplacian}
\end{equation}
Its kernel is isomorphic to the persistent homology group,
\[
    \ker L_q^{a,b}\cong H_q^{a,b},
\]
so the multiplicity of the zero eigenvalue equals the persistent Betti
number $\beta_q^{a,b}$. The positive spectrum contains complementary
nonharmonic information not captured by persistent Betti numbers alone
~\cite{wang2021hermes,memoli2022persistent,jones2025petls}.

The PL spectrum is invariant under filtration-preserving simplicial
relabeling. Reordering or consistently reorienting simplices transforms the
Laplacian by an orthogonal signed-permutation similarity,
\[
    L_q'=P_q^{\top}L_qP_q,
\]
and therefore leaves its spectrum unchanged. For distance-based filtrations,
Euclidean translations and rotations likewise preserve the filtered
simplicial complexes and their Laplacian spectra.

Figure~\ref{fig:math_invariant_example}C illustrates the evolution of the
nonharmonic spectrum of $L_0$ along the filtration. Initially, the complex
contains only isolated vertices and the spectrum is entirely harmonic.
As edges are introduced, positive eigenvalues emerge and evolve with the
connectivity of the simplicial complex. Once the Vietoris--Rips filtration
of the 14-point example reaches the complete clique complex, all positive
eigenvalues coincide.

\subsection{Commutative Algebra (CA)}
\label{sec:persistent_commutative_algebra}

The commutative-algebra (CA) invariant family used in MITNN is based on the persistent Stanley--Reisner theory introduced by Suwayyid and Wei \cite{suwayyid2025persistent}, which associates filtered simplicial complexes with commutative-algebraic objects. This construction encodes the combinatorial organization of a simplicial complex through squarefree monomial ideals, their quotient rings, and facet-associated prime decompositions.

Let $\Delta$ be a finite simplicial complex on the vertex set
$W=\{y_1,\ldots,y_n\}$, and let
$\varphi:\Delta\rightarrow\mathbb{R}$ be a monotone face function satisfying
$\varphi(\omega)\leq\varphi(\omega')$ whenever
$\omega\subseteq\omega'$. The function $\varphi$ induces the filtration
\[
\Delta_\epsilon
=
\{\omega\in\Delta\mid\varphi(\omega)\leq\epsilon\},
\]
with
$\Delta_{\epsilon_a}\subseteq\Delta_{\epsilon_b}$ for
$\epsilon_a\leq\epsilon_b$.
Fix a field $\Bbbk$ and let
$R=\Bbbk[y_1,\ldots,y_n]$ be the standard graded polynomial ring. At each
filtration scale $\epsilon$, the Stanley--Reisner ideal of
$\Delta_\epsilon$ is
\begin{equation}
    I_\epsilon
    =
    I(\Delta_\epsilon)
    =
    \left(
    y_{i_1}\cdots y_{i_s}
    \ \middle|\
    \{y_{i_1},\ldots,y_{i_s}\}\notin\Delta_\epsilon
    \right)
    \subseteq R,
    \label{eq:sr_ideal}
\end{equation}
and the corresponding Stanley--Reisner ring is
$\Bbbk[\Delta_\epsilon]=R/I_\epsilon$. As the filtration expands, the set
of nonfaces decreases, yielding a descending sequence of squarefree
monomial ideals.

The Stanley--Reisner ideal admits an irredundant minimal-prime
decomposition indexed by the facets of the simplicial complex:
\begin{equation}
    I_\epsilon
    =
    \bigcap_{\omega\in\mathcal{F}(\Delta_\epsilon)}
    P_\omega,
    \qquad
    P_\omega
    =
    (y_i\mid y_i\notin\omega),
    \label{eq:facet_prime_decomposition}
\end{equation}
where $\mathcal{F}(\Delta_\epsilon)$ denotes the set of facets of
$\Delta_\epsilon$. To distinguish facets by dimension, define
\[
    \mathcal{P}_\epsilon^{(p)}
    =
    \left\{
    P_\omega
    \ \middle|\
    \omega\in\mathcal{F}(\Delta_\epsilon),
    \ \dim(\omega)=p
    \right\}.
\]
For $\epsilon_a\leq\epsilon_b$, the persistent $p$-dimensional facet set
and its cardinality are
\begin{equation}
    \mathcal{P}_{\epsilon_a,\epsilon_b}^{(p)}
    =
    \mathcal{P}_{\epsilon_a}^{(p)}
    \cap
    \mathcal{P}_{\epsilon_b}^{(p)},
    \qquad
    \mu_p^{\epsilon_a,\epsilon_b}
    =
    \left|
    \mathcal{P}_{\epsilon_a,\epsilon_b}^{(p)}
    \right|.
    \label{eq:facet_persistence_number}
\end{equation}
These quantities characterize the persistence of maximal simplices across
the filtration.

Facet-based quantities describe only maximal simplices, whereas the
$f$-vector records simplices of all dimensions. Let
$\Delta_\epsilon^{(p)}$ denote the set of $p$-simplices in
$\Delta_\epsilon$. The $f$-vector at scale $\epsilon$ is
\begin{equation}
    \mathbf{f}^{\epsilon}
    =
    \left(
    f_0^\epsilon,
    f_1^\epsilon,
    \ldots,
    f_{\dim(\Delta_\epsilon)}^\epsilon
    \right),
    \qquad
    f_p^\epsilon
    =
    \left|
    \Delta_\epsilon^{(p)}
    \right|.
    \label{eq:classical_f_vector}
\end{equation}
Thus, facet persistence and $f$-vectors provide complementary
algebraic--combinatorial information about the evolving simplicial
complex.

The resulting CA quantities are invariant under vertex relabeling. A
permutation of the vertices induces a corresponding permutation of the
polynomial variables and hence a graded automorphism of $R$ that maps the
associated Stanley--Reisner ideals to isomorphic ideals. Consequently, the
facet-associated quantities and $f$-vector components are unchanged.
These quantities therefore define algebraic--combinatorial invariants,
although they are not complete invariants, since nonisomorphic simplicial
complexes may share the same facet counts or $f$-vectors.

Figure~\ref{fig:math_invariant_example}D illustrates the persistence of
$0$-, $1$-, and $2$-dimensional facets along the filtration. A facet is
born when it first appears as a maximal simplex and ceases to persist once
it becomes a face of a higher-dimensional simplex. The corresponding
$P_0$, $P_1$, and $P_2$ bars therefore describe the multiscale evolution
of maximal vertices, edges, and triangles, respectively.

\subsection{Element Interactive Curvature (EIC)}
\label{sec:element_interactive_curvature}

Element interactive curvature (EIC) provides a continuous
differential-geometric representation of element-specific molecular
interactions through differentiable density fields
\cite{nguyen2019dg}. Let a molecular system consist of atoms at positions
$\mathbf{x}_j$ with weights $w_j$. A discrete-to-continuum density field is
defined as
\begin{equation}
    \rho(\mathbf{x};\{\eta_j\},\{w_j\})
    =
    \sum_{j=1}^{N}
    w_j
    \Phi(\|\mathbf{x}-\mathbf{x}_j\|;\eta_j),
    \label{eq:eic_density_reorganized}
\end{equation}
where $w_j=1$ gives a number density and $w_j=q_j$ gives a charge density.
The kernel $\Phi$ is a monotonically decaying $C^2$ radial function
satisfying
\[
    \Phi(d;\eta)\rightarrow1
    \quad\text{as }d\rightarrow0,
    \qquad
    \Phi(d;\eta)\rightarrow0
    \quad\text{as }d\rightarrow\infty.
\]
Common choices include the generalized exponential and generalized Lorentz
kernels,
\begin{equation}
\begin{aligned}
    \Phi_{\mathrm{E}}(d;\eta)
    &=
    \exp\left[-\left(\frac{d}{\eta}\right)^{\kappa}\right],
    && \kappa>0, \\
    \Phi_{\mathrm{L}}(d;\eta)
    &=
    \frac{1}{1+\left(\dfrac{d}{\eta}\right)^{\nu}},
    && \nu>0 .
\end{aligned}
\label{eq:eic_correlation_kernels}
\end{equation}

To retain chemical specificity, the density construction is resolved by
element pairs. Let
$    D_a
    =
    \bigcup_{i:e_i=a} B(\mathbf{x}_i,r_i)$
denote the van der Waals domain associated with atoms of element type $a$.
For two distinct element types $a$ and $b$, the element-interactive density
generated by atoms of type $b$ over $D_a$ is
\begin{equation}
    \rho_{ab}(\mathbf{x};\eta_{ab})
    =
    \sum_{\substack{j:e_j=b\\
    \|\mathbf{x}_i-\mathbf{x}_j\|>
    r_i+r_j+\sigma,\ \forall\, i:e_i=a}}
    w_j
    \Phi(\|\mathbf{x}-\mathbf{x}_j\|;\eta_{ab}),
    \qquad
    \mathbf{x}\in D_a,
    \label{eq:eic_element_density}
\end{equation}
where $r_i$ and $r_j$ are atomic radii and $\sigma$ is the
dataset-dependent correction used in the original EIC construction.
The distance constraint excludes covalent interactions. Interactions
between atoms of the same element type are treated using atom-specific
van der Waals domains to avoid self-contributions.

A multiscale representation is obtained by varying the characteristic
interaction length. Following the EIC parametrization,
\[
    \eta_{ab}(\tau)
    =
    \tau\bigl(\bar r_a+\bar r_b\bigr),
\]
where $\bar r_a$ and $\bar r_b$ are the van der Waals radii associated
with element types $a$ and $b$, and $\tau$ is a dimensionless scale
parameter. Different values of $\tau$ probe element-specific interactions
over different spatial ranges.

For a fixed element pair and interaction scale, element-interactive
manifolds are defined by the level sets
\[
    \rho_{ab}(\mathbf{x};\eta_{ab})
    =
    c\,\rho_{\max},
    \qquad
    0\leq c\leq1,
\]
where
 $   \rho_{\max}
    =
    \max_{\mathbf{x}\in D_a}
    \rho_{ab}(\mathbf{x};\eta_{ab}).$
At regular points for which $\nabla\rho\neq0$, let
$    g=\rho_x^2+\rho_y^2+\rho_z^2.$
The Gaussian curvature $K$ and mean curvature $H$ are obtained from the
first- and second-order spatial derivatives of the density:
\begin{align}
    K
    =&
    \frac{1}{g^2}
    \Big[
    2\rho_x\rho_y\rho_{xz}\rho_{yz}
    +2\rho_x\rho_z\rho_{xy}\rho_{yz}
    +2\rho_y\rho_z\rho_{xy}\rho_{xz}
    \nonumber\\
    &\quad
    -2\rho_x\rho_z\rho_{xz}\rho_{yy}
    -2\rho_y\rho_z\rho_{xx}\rho_{yz}
    -2\rho_x\rho_y\rho_{xy}\rho_{zz}
    \nonumber\\
    &\quad
    +\rho_z^2\rho_{xx}\rho_{yy}
    +\rho_x^2\rho_{yy}\rho_{zz}
    +\rho_y^2\rho_{xx}\rho_{zz}
    -\rho_x^2\rho_{yz}^2
    -\rho_y^2\rho_{xz}^2
    -\rho_z^2\rho_{xy}^2
    \Big],
    \label{eq:eic_gaussian_curvature}
\end{align}
and
\begin{align}
    H
    =&
    \frac{1}{2g^{3/2}}
    \Big[
    2\rho_x\rho_y\rho_{xy}
    +2\rho_x\rho_z\rho_{xz}
    +2\rho_y\rho_z\rho_{yz}
    \nonumber\\
    &\quad
    -(\rho_y^2+\rho_z^2)\rho_{xx}
    -(\rho_x^2+\rho_z^2)\rho_{yy}
    -(\rho_x^2+\rho_y^2)\rho_{zz}
    \Big].
    \label{eq:eic_mean_curvature_reorganized}
\end{align}
The corresponding principal curvatures are
\[
    \kappa_{\min}
    =
    H-\sqrt{H^2-K},
    \qquad
    \kappa_{\max}
    =
    H+\sqrt{H^2-K}.
\]
These curvature functions characterize the local geometry associated with
each element pair and interaction scale.

Because the density construction depends on radial functions of Euclidean
distances, atom reindexing leaves it unchanged, while translations and
rotations rigidly transform the density field and its level sets without
changing their curvature values. Gaussian curvature is independent of
surface orientation, whereas the sign of mean curvature depends on the
chosen normal direction. Under a consistent normal convention, the EIC
quantities are therefore invariant under atom reindexing, translations,
and rotations.

\subsection{Forman Persistent Ricci Curvature (FPRC)}
\label{sec:forman_persistent_ricci_curvature}

Forman persistent Ricci curvature (FPRC) characterizes the evolution of
discrete combinatorial curvature across a filtration of simplicial complexes
~\cite{wee2021forman}. Two distinct $p$-simplices $\alpha$ and
$\bar{\alpha}$ are called parallel, denoted by
$\bar{\alpha}\parallel\alpha$, if they share a common $(p-1)$-face but
are not both faces of any common $(p+1)$-simplex.

For $p>0$, the Forman Ricci curvature of a $p$-simplex $\alpha$ in an
unweighted simplicial complex is
\begin{equation}
    F_p(\alpha)
    =
    \#\{\beta^{p+1}\mid \alpha<\beta^{p+1}\}
    +
    \#\{\gamma^{p-1}\mid \gamma^{p-1}<\alpha\}
    -
    \#\{\bar{\alpha}\mid \bar{\alpha}\parallel\alpha\},
    \label{eq:fprc_frc_reorganized}
\end{equation}
where the first term counts the $(p+1)$-dimensional cofaces of $\alpha$,
the second counts its $(p-1)$-dimensional faces, and the third counts its
parallel $p$-simplices. 

Following the convention used in the FPRC construction
~\cite{wee2021forman}, the curvature of a nonisolated vertex $v$ is
defined from the curvatures of its incident edges as
\begin{equation}
    F_0(v)
    =
    \frac{1}{d(v)}
    \sum_{e>v} F_1(e),
    \label{eq:fprc_vertex}
\end{equation}
where $d(v)$ is the number of edges incident to $v$. For an isolated
vertex, we set $F_0(v)=0$. Evaluating Forman Ricci curvature across simplex
dimensions and filtration values yields a multiscale characterization of
the evolving combinatorial curvature.

Forman Ricci curvature depends only on the local face--coface incidence
relations and parallel-simplex relations. These relations are preserved under
simplicial isomorphisms and vertex relabeling, so corresponding simplices
have identical curvature values. For distance-based filtrations, Euclidean
translations and rotations also preserve the filtered simplicial complexes.
The resulting FPRC quantities are therefore invariant under these
transformations.

Figure~\ref{fig:math_invariant_example}E illustrates the signed mean Forman
curvature $\operatorname{mean}(F_p)$ for $p=0,1,2$. Isolated vertices are
assigned $F_0=0$. The first edge has $F_1=2$, whereas the first triangle
has $F_2=3$, reflecting their respective local face--coface configurations
at the time of appearance. At intermediate filtration values, the mean
curvature varies as the local face--coface and parallel-simplex
configurations evolve. For the 14-point example, the three mean-curvature
curves eventually converge and remain constant once the Vietoris--Rips filtration reaches the complete clique complex.

 \section*{Acknowledgments}
 This work was supported in part by NIH grant  R35GM148196, the University of Georgia,   the Georgia Research Alliance and the Defense Advanced Research Projects Agency (DARPA) under Agreement No. HR00112969E087.
 MH was supported in part by the National Science Foundation (NSF grant DMS-2134231).
The views and conclusions contained in this document are those of the authors and should not be interpreted as representing the official policies, either expressed or implied, of the U.S. Government.
 
\bibliographystyle{unsrtnat}
\bibliography{reference}
\end{document}